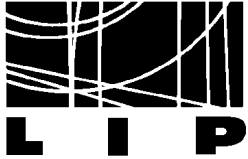

# Resistive plate chambers for time-of-flight measurements

A.Blanco[1], P.Fonte[1,2,*], L.Lopes[1,3], A.Mangiarotti[4],
R. Ferreira-Marques[1,3], A. Policarpo[1,3]

1 - LIP – Laboratório de Instrumentação e Física Experimental de Partículas, Portugal
2 - Instituto Superior de Engenharia de Coimbra, Coimbra, Portugal
3 - CFRM, Departamento de Física da Universidade de Coimbra, Coimbra, Portugal.
4 - Università degli Studi di Firenze, Dipartimento di Fisica, Largo E.Fermi 2, 50125 Firenze, Italy

## Abstract

The applications of Resistive Plate Chambers (RPCs) have recently been extended by the development of counters with time resolution below 100 ps s for minimum ionising particles. Applications to HEP experiments have already taken place and many further applications are under study.

In this work we address the operating principles of such counters along with some present challenges, with emphasis on counter aging.

[*] Corresponding author. LIP - Coimbra, Departamento de Física da Universidade de Coimbra, 3004-516 Coimbra, PORTUGAL. Tel: (+351) 239 833 465, fax: (+351) 239 822 358, e-mail: fonte@lipc.fis.uc.pt

# 1. Introduction

Resistive Plate Chambers (RPCs) with time resolution below 100 ps σ for minimum ionising particles have been recently developed [1], [2]. This type of detector, operating at atmospheric pressure with non-flammable gases, seems well suited for high-granularity time-of-flight (TOF) systems, providing performances comparable to the scintillator-based TOF technology but offering a significantly lower price per channel, compact mechanics and magnetic field compatibility. A recent review may be found in [3].

Time resolutions of 300 ps FWHM have been also demonstrated for 511 keV photon pairs, with possible application to whole-body human PET imaging. The sub-millimetric imaging of small animals may be also an attractive possibility [4].

# 2. Some considerations on the working principles

## 2.1. Efficiency

It has been recognized long ago that the large efficiency values reached by RPCs for minimum ionising particles (MIPs) cannot be explained without resorting to some form of avalanche growth saturation, possibly due to the space charge effect [5], which extends the sensitive region (close to the cathode).

The width of the sensitive region may be estimated by requiring that the measured inefficiency $(1-\varepsilon)$ equals the probability that a passing particle deposits no ionisation in it. The required average number of "effective" ionisation clusters is then

$$N^* = -\ln(1-\varepsilon) \qquad 1)$$

and the width of the sensitive region $g^* = N^*/\lambda$, where $\lambda$ is the cluster density per unit length.

Figure 1 shows some data on timing RPC efficiency for different gases and gap widths from which $N^*$ and the sensitive fraction of the gas gap, $g^*/g$, may be calculated. In all cases the sensitive region corresponds to about half of the gap width.

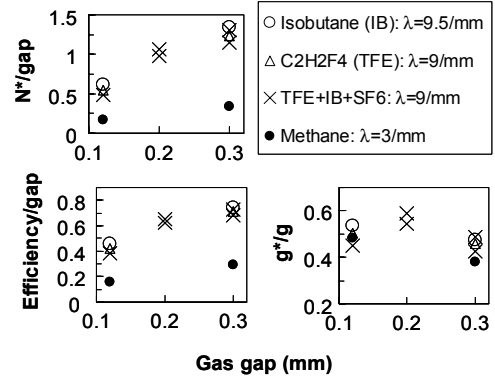

Figure 1 – Efficiency data taken from [3] for different gases and gap widths ($g$) along with the corresponding sensitive fraction of the gas gap ($g^*/g$) and the average effective number of primary clusters ($N^*$) per gap. Data for λ was taken from [8] (highly relativistic particles).

## 2.2. Timing Resolution

A simple theory of the RPC timing properties may be formulated on the following principles (see also [6], [7], [8]).

The poissonian probability that $k>0$ clusters will be produced in the sensitive region of the gap from the $N^*$ average effective clusters is given by

$$P(k) = \frac{(N^*)^k}{(\exp(N^*)-1)k!}. \qquad 2)$$

Avalanches initiated by (in first approximation) single electrons will result in an exponentially distributed final avalanche charge (Furry law [9]) after a fixed development length[1]. Since the gain fluctuations occur mainly in the first few amplification steps (see [8] for a detailed study) the effective initial number of electrons, $k$, may be replaced by an apparent value, $n$, that takes into account both the cluster and the avalanche statistics,

$$n = \sum_1^k h, \qquad 3)$$

---

[1] Calculations suggest that the detailed form of the avalanche gain statistics has a negligible influence on the time resolution ([6], [8]), being reasonable to consider the simplest case.

where $h$ is an exponentially distributed random variable with unit mean value.

The distribution of $n$ may be calculated analytically [10], yielding

$$P(n) = e^{-n} \frac{N^* I_1(2\sqrt{nN^*})}{(\exp(N^*)-1)\sqrt{nN^*}}, \qquad 4)$$

where $I_1$ is the modified Bessel function of the first kind. In the limit of very small $N^*$ the probability of releasing more than one cluster vanishes and $P(n) \to e^{-n}$, recovering the Furry law.

- The $n$ apparent initial charges are deterministically multiplied in time by a factor $\exp(\alpha^* v t)$, where $\alpha^*$ is the effective multiplication coefficient (first Townsend coefficient minus the electrons attachment coefficient), $v$ the electron drift velocity and $t$ the elapsed time. The induced current is therefore given by $i(t) = n\, i_e \exp(\alpha^* v t)$, with $i_e$ being the current induced by a single drifting electron.
- Owing to the exponential growth of current in time the comparator will sense also an exponentially growing voltage with the same value of $\alpha^* v$ [11]. Therefore the threshold voltage maybe related to an equivalent threshold current $i_{th}$ at the amplifier input.
- It is assumed that the space charge effect will become important only for induced currents larger than $i_{th}$.
- When the threshold current is reached, a time $T$ is registered in accordance with

$$i_{th} = n\, i_e \exp(\alpha^* v T) = n\, i_e \exp(\tau). \qquad 5)$$

where the reduced variable $\tau$ was introduced, corresponding to the measured time in units of $1/(\alpha^* v)$.

Eq. 5) relates functionally the reduced measured time $\tau$ to the distribution 4) and therefore the distribution of $\tau$ may be calculated by a change of variable. Restating 5) as $\exp(\tau') = n \exp(\tau)$ one finds

$$P(\tau) = e^{(\tau'-\tau)-\exp(\tau'-\tau)} \frac{N^* I_1(2\sqrt{N^* e^{(\tau'-\tau)}})}{(\exp(N^*)-1)\sqrt{N^* e^{(\tau'-\tau)}}}. \qquad 6)$$

In the single cluster limit ($N^* \to 0$)

$$P(\tau) \to e^{(\tau'-\tau)-\exp(\tau'-\tau)}, \qquad 7)$$

reproducing the result derived in [6].

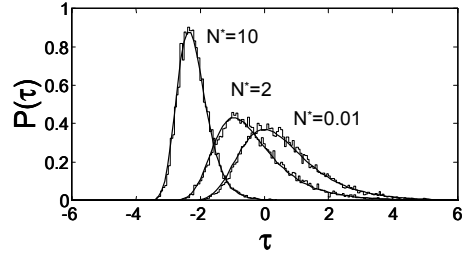

Figure 2 – Distribution, from eq.6), of the reduced measured time $\tau = \alpha^* v T$ (thick line), for different values of $N^*$. The superimposed histograms correspond to an equivalent Monte-Carlo calculation.

A graphical representation of eq.6) is shown in Figure 2, along with an equivalent Monte-Carlo calculation adapted from [12]. However, the Monte-Carlo is more accurate in that all clusters are followed to the anode and $N^*$ is determined from the simulation output. The good agreement shows that the "effective number of primary clusters" approach is highly accurate.

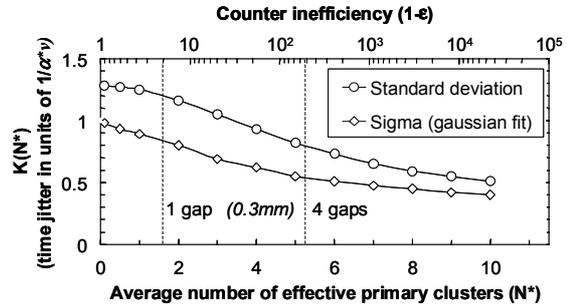

Figure 3 – Width of $P(\tau)$ (eq.6)) evaluated by its standard deviation $K_2(N^*)$ (eq.8)), following [7], and by fitting a gaussian curve (the method used to analyse experimental data). The expected values for $N^*$ (from Figure 1) are marked for 1 and 4 (0.3 mm) gap counters.

Since any change on the threshold level affects only $\tau'$ it is clear from the functional form of eq.6) that this results merely on a shift of the time distribution. Therefore any moment about the mean will depend only on $N^*$ (or, equivalently, on the counter inefficiency) and can be written, going back to the measured time $T$,

$$E[(T-E(T))^m] = \frac{K_m(N^*)}{\alpha^* v} = \frac{K_m(-\ln(1-\varepsilon))}{\alpha^* v} \qquad 8)$$

The reduced standard deviation $K_2(N^*)$ may be calculated as a series [7], being the results shown in Figure 3 along with a gaussian fit to $P(\tau)$ (eq.6).

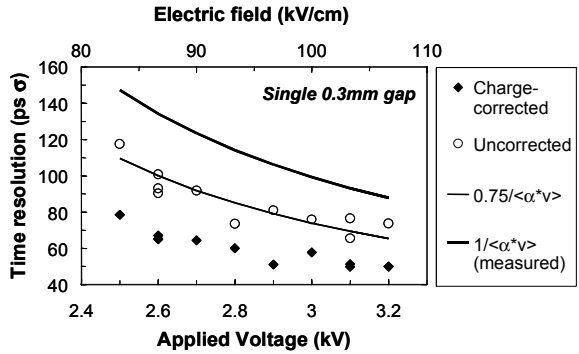

Figure 4 – Comparison, based on eq.8), between the measured average values of $\alpha^* v$ (lines) and the counter time resolution (with and without charge correction). A reasonable value of $K(N^*)=0.75$ reproduces well the uncorrected data (compare with Figure 3).

Presented in Figure 4 is a comparison between experimental data and the present model. The average value of $\alpha^* v$ was determined using the method described in [11] along with the time resolution data, using the setup described in [13]. The theory reasonably describes the data when no charge correction is applied to the measured time. This is natural, since the time-charge correlation was not considered in the model.

The present treatment expands upon previous work [7] in that an explicit expression for $P(\tau)$ is obtained. This allows the measurement of the distribution width by a gaussian adjustment, providing a more accurate comparison with the experimentally available data.

## 3. Present challenges

A number of challenges still stand on the way of full-scale applications of high resolution TOF with RPCs. Some of the most important ones include:
- the demonstration that a high-granularity RPC-TOF counter is able to operate without loss of time resolution in a high occupancy environment despite potentially large crosstalk levels;
- investigation of the aging effects (see below).

For many applications it would be also of fundamental importance the extension of the counting rate capabilities from around 200 Hz/cm$^2$ (for most glass-made counters) to much larger values and the development of effective methods for controlling the timing tails (see [13] for a possible approach).

### 3.1. Aging studies

Severe aging of glass RPCs operating in streamer mode has been observed and related to the presence of water vapour traces in the gas mixture. An unidentified deposit was found over the glass surface, severely increasing the dark count rates and reducing the counter efficiency [14].

Naturally it is of great practical importance the investigation of such effect in timing RPCs, often made with glass electrodes and working in somewhat similar gaseous mixtures. To this aim we built a test setup comprising six single-gap counters, each made with one glass and one aluminium electrode. Three counters were connected with the glass electrode as a cathode and three were connected with the glass electrode as an anode.

All counters were illuminated laterally, along the gas gap, by a mercury UV lamp that created primary charges by photoelectric effect at the cathodes. The different cathode photosensitivities were compensated by slightly adjusting the applied voltages, ranging from 2.8 to 3.0 kV, until all chambers registered similar current values. For the settings chosen the currents were essentially independent from any small variations in the lamp intensity, having reached the maximum value allowed by the resistive electrode.

The counters were illuminated for 22 hours and kept in darkness 2 hours per day, being continuously recorded the counter currents, temperature, atmospheric pressure and lamp intensity. The setup was kept in a continuous flow of "standard" timing RPC gas, $C_2H_2F_4+10\%SF_6+5\%iso\text{-}C_4H_{10}$, being added enough water vapour to keep a 10% relative humidity. The gas container was made of acrylic plastic with Viton o-ring seals and several common types of glues, plastics and metals were used internally as needed.

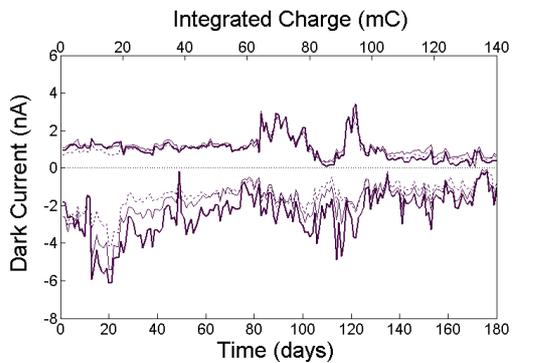

Figure 5 – Dark current as a function of time and accumulated charge for 6 counters (positive: glass cathode, negative: aluminium cathode). No systematic increase in dark current is visible, contrasting with the aging effect observed in streamer-mode RPCs [14]. The integrated charge corresponds to 800 days of operation at the typical counter operational settings.

The dark current values recorded daily at the end of the 2-hour dark period were corrected for systematic variations due to temperature, being the results for six months of operation shown in Figure 5. There is no evidence of any systematic long-term increase of dark current, suggesting that the aging effect mentioned above is less severe in timing RPCs than in streamer-mode RPCs. The short-term dark current fluctuations are well correlated for the three chambers of the same type (glass or aluminium cathode) and therefore should be of environmental origin.

However, a visual inspection of the electrodes confirmed the existence of a whitish, dry, deposit over the glass cathodes. The deposit covered an area of approximately 1 cm$^2$ close to the UV-light entrance slit, presumably corresponding to the effectively illuminated region. The glass anodes were also thinly covered with a viscous-looking substance everywhere excepted in the 1 cm$^2$ irradiated region. The aluminium electrodes were perfectly clear.

The average integrated charge of 140 mC (changes slightly from chamber to chamber) corresponds to 800 days of operation at 200 Hz/cm$^2$ and 10 pC/avalanche.

## 4. Conclusions

Experimental efficiency data shows that timing RPCs are sensitive to primary ionisation clusters created in about half of the gas gap (cathode-side). For 0.3 mm-gap devices filled with common gases the average number of primary clusters created in this region ($N^*$) is close to 1/gap for MIPs.

A simple, analytically solvable, model of the time resolution suggests that it should depend only on some function $K(N^*)$ divided by $\alpha^* v$. Multiple gaps are taken in consideration simply through the value of $N^*$. The model reproduces well the available experimental data when the charge-time correlation is not considered.

Aging studies were performed on six counters made with glass and aluminium electrodes. After a charge transfer equivalent to 800 days of normal operation no increase of dark current was detected. However an unidentified deposit was found over the glass cathodes, being absent on the irradiated regions of the glass anodes and on all aluminium electrodes.

### Acknowledgments

The competent technical work of Nuno Carolino and Américo Pereira is gratefully acknowledged.

This work was financed by "Fundação para a Ciência e Tecnologia" project CERN/FNU/43723/2001.